\documentclass[review,12pt]{elsarticle}

\usepackage{graphicx,amssymb}
\usepackage{geometry}
\usepackage{dcolumn}
\usepackage[tbtags]{amsmath}

\newcommand{\zh}{z}

\newcommand{\xbj}{x}
\newcommand{\ph}{\phi_h}

\begin{document}

\begin{frontmatter}

\title{{\large  Precise Measurements of Beam Spin Asymmetries in Semi-Inclusive\\
$\pi^0$ production}}

%%%%%%%%%%%%%%%%%%%%%%%%%%%%%%%%%%% 
% FOLLOWING A GENERIC TAIL FOR TEX
%   Nothing here relates to author list
%   Suggestion for acknowlegements.  Could be deleted.
%%%%%%
\newcommand* {\ANL}{Argonne National Laboratory, Argonne, Illinois 60441}
\newcommand* {\ANLindex}{1}
\newcommand* {\ASU}{Arizona State University, Tempe, Arizona 85287-1504}
\newcommand* {\ASUindex}{2}
\newcommand* {\UCLA}{University of California at Los Angeles, Los Angeles, California  90095-1547}
\newcommand* {\UCLAindex}{3}
\newcommand* {\CSUDH}{California State University, Dominguez Hills, Carson, CA 90747}
\newcommand* {\CSUDHindex}{4}
\newcommand* {\CANISIUS}{Canisius College, Buffalo, NY}
\newcommand* {\CANISIUSindex}{5}
\newcommand* {\CMU}{Carnegie Mellon University, Pittsburgh, Pennsylvania 15213}
\newcommand* {\CMUindex}{6}
\newcommand* {\CUA}{Catholic University of America, Washington, D.C. 20064}
\newcommand* {\CUAindex}{7}
\newcommand* {\SACLAY}{CEA, Centre de Saclay, Irfu/Service de Physique Nucl\'eaire, 91191 Gif-sur-Yvette, France}
\newcommand* {\SACLAYindex}{8}
\newcommand* {\CNU}{Christopher Newport University, Newport News, Virginia 23606}
\newcommand* {\CNUindex}{9}
\newcommand* {\UCONN}{University of Connecticut, Storrs, Connecticut 06269}
\newcommand* {\UCONNindex}{10}
\newcommand* {\EDINBURGH}{Edinburgh University, Edinburgh EH9 3JZ, United Kingdom}
\newcommand* {\EDINBURGHindex}{11}
\newcommand* {\FU}{Fairfield University, Fairfield CT 06824}
\newcommand* {\FUindex}{12}
\newcommand* {\FIU}{Florida International University, Miami, Florida 33199}
\newcommand* {\FIUindex}{13}
\newcommand* {\FSU}{Florida State University, Tallahassee, Florida 32306}
\newcommand* {\FSUindex}{14}
\newcommand* {\Genova}{Universit$\grave{a}$ di Genova, 16146 Genova, Italy}
\newcommand* {\Genovaindex}{15}
\newcommand* {\GWUI}{The George Washington University, Washington, DC 20052}
\newcommand* {\GWUIindex}{16}
\newcommand* {\ISU}{Idaho State University, Pocatello, Idaho 83209}
\newcommand* {\ISUindex}{17}
\newcommand* {\INFNFE}{INFN, Sezione di Ferrara, 44100 Ferrara, Italy}
\newcommand* {\INFNFEindex}{18}
\newcommand* {\INFNFR}{INFN, Laboratori Nazionali di Frascati, 00044 Frascati, Italy}
\newcommand* {\INFNFRindex}{19}
\newcommand* {\INFNGE}{INFN, Sezione di Genova, 16146 Genova, Italy}
\newcommand* {\INFNGEindex}{20}
\newcommand* {\INFNRO}{INFN, Sezione di Roma Tor Vergata, 00133 Rome, Italy}
\newcommand* {\INFNROindex}{21}
\newcommand* {\ORSAY}{Institut de Physique Nucl\'eaire ORSAY, Orsay, France}
\newcommand* {\ORSAYindex}{22}
\newcommand* {\ITEP}{Institute of Theoretical and Experimental Physics, Moscow, 117259, Russia}
\newcommand* {\ITEPindex}{23}
\newcommand* {\JMU}{James Madison University, Harrisonburg, Virginia 22807}
\newcommand* {\JMUindex}{24}
\newcommand* {\KNU}{Kyungpook National University, Daegu 702-701, Republic of Korea}
\newcommand* {\KNUindex}{25}
\newcommand* {\LPSC}{LPSC, Universite Joseph Fourier, CNRS/IN2P3, INPG, Grenoble, France
}
\newcommand* {\LPSCindex}{26}
\newcommand* {\UNH}{University of New Hampshire, Durham, New Hampshire 03824-3568}
\newcommand* {\UNHindex}{27}
\newcommand* {\NSU}{Norfolk State University, Norfolk, Virginia 23504}
\newcommand* {\NSUindex}{28}
\newcommand* {\OHIOU}{Ohio University, Athens, Ohio  45701}
\newcommand* {\OHIOUindex}{29}
\newcommand* {\ODU}{Old Dominion University, Norfolk, Virginia 23529}
\newcommand* {\ODUindex}{30}
\newcommand* {\RPI}{Rensselaer Polytechnic Institute, Troy, New York 12180-3590}
\newcommand* {\RPIindex}{31}
\newcommand* {\URICH}{University of Richmond, Richmond, Virginia 23173}
\newcommand* {\URICHindex}{32}
\newcommand* {\ROMAII}{Universita' di Roma Tor Vergata, 00133 Rome Italy}
\newcommand* {\ROMAIIindex}{33}
\newcommand* {\MSU}{Skobeltsyn Nuclear Physics Institute, Skobeltsyn Nuclear Physics Institute, 119899 Moscow, Russia}
\newcommand* {\MSUindex}{34}
\newcommand* {\SCAROLINA}{University of South Carolina, Columbia, South Carolina 29208}
\newcommand* {\SCAROLINAindex}{35}
\newcommand* {\JLAB}{Thomas Jefferson National Accelerator Facility, Newport News, Virginia 23606}
\newcommand* {\JLABindex}{36}
\newcommand* {\UTFSM}{Universidad T\'{e}cnica Federico Santa Mar\'{i}a, Casilla 110-V Valpara\'{i}so, Chile}
\newcommand* {\UTFSMindex}{37}
\newcommand* {\GLASGOW}{University of Glasgow, Glasgow G12 8QQ, United Kingdom}
\newcommand* {\GLASGOWindex}{38}
\newcommand* {\VT}{Virginia Polytechnic Institute and State University, Blacksburg, Virginia   24061-0435}
\newcommand* {\VTindex}{39}
\newcommand* {\VIRGINIA}{University of Virginia, Charlottesville, Virginia 22901}
\newcommand* {\VIRGINIAindex}{40}
\newcommand* {\WM}{College of William and Mary, Williamsburg, Virginia 23187-8795}
\newcommand* {\WMindex}{41}
\newcommand* {\YEREVAN}{Yerevan Physics Institute, 375036 Yerevan, Armenia}
\newcommand* {\YEREVANindex}{42}

\newcommand* {\NOWLANL}{Los Alamos National Laborotory, New Mexico, NM}
\newcommand* {\NOWMSU}{Skobeltsyn Nuclear Physics Institute, Skobeltsyn Nuclear Physics Institute, 119899 Moscow, Russia}
\newcommand* {\NOWINFNGE}{INFN, Sezione di Genova, 16146 Genova, Italy}
\newcommand* {\NOWJLAB}{Thomas Jefferson National Accelerator Facility, Newport News, Virginia 23606}
 %%%%%%%%%%%%%%% END OF Latex Macros for institute addresses  %%%%%%%%%%%%%%%%%%%%%%%%% 

\author[toINFNFR]{M.~Aghasyan }
\ead{mher@jlab.org}
\author[toJLAB]{H.~Avakian}
\author[toINFNFR]{P.~Rossi}
 \author[toINFNFR]{E.~De Sanctis}
\author[toINFNFR]{D.~Hasch}
\author[toINFNFR]{M.~Mirazita}
\author[toODU]{D.~Adikaram}
\author[toODU]{M.J.~Amaryan}
\author[toINFNGE]{M.~Anghinolfi}
\author[toVIRGINIA,toODU]{H.~Baghdasaryan}

% ...
\address[toINFNFR]{\INFNFR} 
 \address[toINFNGE]{\INFNGE} 
% ...
\address[toODU]{\ODU} 
% ...
\address[toJLAB]{\JLAB} 
% ...
\address[toVIRGINIA]{\VIRGINIA} 
% ...

%%%%%%
\date{\today}
\begin{abstract}

We present studies of single-spin asymmetries for neutral pion electroproduction in semi-inclusive 
deep-inelastic scattering of 5.776 GeV polarized electrons from
an unpolarized hydrogen target, 
 using the CEBAF 
Large Acceptance 
Spectrometer (CLAS)
at the Thomas Jefferson National Accelerator
Facility.  
A substantial $\sin \phi_h$ amplitude has been measured 
in the distribution of the cross section asymmetry 
as a function of the azimuthal angle $\phi_h$ of the produced neutral
pion. 
The dependence of 
this amplitude on Bjorken $\xbj$ and on the pion transverse
momentum is extracted with significantly higher precision than previous data and is compared to model calculations.  

\end{abstract}
%\maketitle
\end{frontmatter}

In recent years it has become clear that understanding the orbital motion of partons
is crucial 
for achieving a more complete picture of the nucleon in terms of elementary quarks 
and gluons. 
Parton distribution functions have been generalized to contain information not only on the 
longitudinal momentum but also on  the transverse momentum distributions of partons in a fast moving hadron.
Intense theoretical investigations of Transverse Momentum Dependent (TMD) distributions of partons
and the first unambiguous experimental signals of TMDs
indicate that QCD-dynamics inside hadrons is  
much richer than what can be learned from collinear parton distributions.

TMDs were first suggested to explain the large transverse single-spin asymmetries observed
in polarized hadron-hadron collisions.
Since then, two fundamental mechanisms involving transverse momentum dependent distributions and/or
fragmentation functions have been identified, which lead to  
 single-spin asymmetries (SSAs) in hard processes:
a) internal quark motion as represented by, e.g., the Sivers mechanism
~\cite{Sivers:1990fh,Anselmino:1998yz,Brodsky:2002cx,Collins:2002kn,Ji:2002aa}, 
which 
generates an asymmetric distribution of quarks in a nucleon that is transversely polarized
and 
b) the Collins mechanism \cite{Collins:2002kn,Mulders:1995dh}, which correlates the transverse spin
of the struck quark with the transverse momentum of the observed hadron. 
The 'Sivers-type' mechanism requires non-zero orbital angular momentum of the struck parton
together with initial- or final-state interactions via soft-gluon 
exchange~\cite{Brodsky:2002cx,Collins:2002kn,Ji:2002aa}.
This mechanism involves TMD distributions which describe the correlations between
the  transverse motion of the parton and its own transverse spin or
the spin of the initial- or final-state hadron, thereby providing unprecedented information 
about spin-orbit correlations.

Semi-inclusive deep-inelastic scattering (SIDIS) has emerged as a powerful tool to probe 
nucleon structure and to provide access to TMDs through measurements of spin and azimuthal asymmetries.
A rigorous basis for such studies of TMDs in SIDIS is provided by
TMD factorization in QCD, which has been established in 
Refs.~\cite{Ji:2004wu,Collins:2004nx, Bacchetta:2008xw} for 
leading twist\footnote{each twist increment above leading twist (twist-2) contributes an extra suppression factor of $1/Q$}
single hadron production with transverse momenta being much smaller than the hard scattering scale.
In this kinematic domain, the SIDIS cross section 
can be expressed in terms of 
structure functions 
\cite{Mulders:1995dh,Levelt:1994np,Bacchetta:2006tn} which are certain convolutions of 
transverse momentum dependent distribution and fragmentation
functions. 
The analysis of TMDs thus strongly depends on the knowledge of fragmentation functions~\cite{PhysRevD.75.114010,Amrath:2005gv,Bacchetta:2007wc,PhysRevD.83.074003,PhysRevD.75.094009}.

Many different observables, which help to pin down various TMD effects, are currently 
available from experiments such as:
1) semi-inclusive deep-in-elastic scattering 
(HERMES at DESY~\cite{Airapetian:1999tv,Airapetian:2001eg,Airapetian:2004tw,Airapetian:2006rx,Airapetian:2009ti,Airapetian:2010ds}, 
COMPASS 
at CERN~\cite{Alekseev:2010rw,Alexakhin:2005iw, Collaboration:2010fi},  and Jefferson 
Lab~\cite{Avakian:2003pk,Avakian:2005ps,Avakian:2010ae,Mkrtchyan:2007sr}), 
2) polarized proton-proton collisions (BRAHMS, PHENIX and STAR at RHIC
 \cite{Adams:2003fx,Chiu:2007zy,Arsene:2008mi,Adler:2005in,:2008qb,Adare:2010bd}) 
and 
3) electron-positron annihilation (Belle at KEK~\cite{Abe:2005zx,Seidl:2008xc}).

This letter reports measurements of single-spin asymmetries in the production
of neutral pions 
by longitudinally polarized electrons scattered off unpolarized protons.
The helicity-dependent part ($\sigma_{LU}$) 
arises from the anti-symmetric part of the hadronic tensor \cite{Bacchetta:2006tn}:

\begin{eqnarray}
\label{HLT}
%\begin{split}
\frac{d\sigma_{LU}}{d\xbj dy\,d\zh dP^2_T d\ph}  =\frac{2\pi \alpha^2}{\xbj y Q^2}\frac{y^2}{2(1-\varepsilon)}
   \times \,\, \nonumber   \\ 
    \Bigl( 1+ \frac{\gamma^2}{2\xbj} \Bigr) \lambda_e \sqrt{2\varepsilon(1+\varepsilon)}  \sin \phi_h \,\,F^{\sin \ph}_{LU},
\end{eqnarray}
 
\noindent  with the structure function:
 \begin{eqnarray}
\label{FLUf}
   F^{\sin \ph}_{LU} =  \frac{2M}{Q} \,\,% \nonumber   \\ 
   \int d^2 \! \boldsymbol{p_T} d^2 \boldsymbol{k_T} \, \times \,\, \nonumber \\
         \delta^{(2)}
          \Bigl(\boldsymbol{p_T}
         -\frac{\boldsymbol{P_{T}}}{z}-\boldsymbol{k_T}\Bigr) \times \,\,  \nonumber \\
   \biggl\{ \frac{\boldsymbol{\hat{P}_{T}}\cdot
         \boldsymbol{p_T}}{M}
         \left[\frac{M_h}{M}\, h_1^{\perp} \frac{\tilde{E}}{z} +\xbj\, g^\perp D_1\right] -  \nonumber \\
    \frac{\boldsymbol{\hat{P}_{T}}
         \cdot \boldsymbol{k_T}}{M_h}
         \left[\frac{M_h}{M}\,f_1\, \frac{\tilde{G^\perp }}{z} + \xbj\, e H_1^{\perp} \right] \biggr\}. 
  %\end{split}
\end{eqnarray}

\noindent The subscripts $LU$ specify the beam and target 
polarizations ($L$ stands for longitudinally 
polarized and $U$ for unpolarized), $\alpha$ is the fine structure constant and
 $\ph$ is the  azimuthal angle between the leptonic and the hadronic 
planes defined according to the Trento convention \cite{Bacchetta:2004jz}.
The kinematic variables $\xbj$, $y$, and $z$  are defined as: 
$\xbj = Q^2/{2(\boldsymbol{P}_1 \cdot \boldsymbol{q})}$, $y={(\boldsymbol{P}_1 \cdot \boldsymbol{q})/(\boldsymbol{P}_1 \cdot \boldsymbol{k}_1)}$, $\zh=(\boldsymbol{P}_1 \cdot \boldsymbol{P})/(\boldsymbol{P}_1 \cdot \boldsymbol{q})$, 
where $Q^2=-q^2=-(\boldsymbol{k}_1-\boldsymbol{k}_2)^2$ is the four-momentum 
of the virtual photon, $\boldsymbol{k}_1$ ($\boldsymbol{k}_2$) is the four-momentum of the incoming (scattered) lepton,
$\boldsymbol{P}_1$ and $\boldsymbol{P}$ are the four-momenta of the target nucleon and the observed final-state 
hadron, respectively, $\lambda_e$ is the electron beam helicity, $\gamma=2M\xbj /Q$,
 $M$ and $M_h$ are the nucleon and hadron masses, $\boldsymbol{P_{T}}$ is the transverse momentum of the 
detected hadron (with $\boldsymbol{\hat{P}_{T}}=\boldsymbol{P_{T}}/|\boldsymbol{P_{T}}|$), and
$\boldsymbol{p_T}$ and $\boldsymbol{k_T}$ are the intrinsic quark transverse momenta in the 
distribution function (DF) and fragmentation function (FF), respectively. In Eq.~\ref{FLUf}
 we use small and capital letters for DF and FF, respectively. The ratio $\varepsilon$ of the longitudinal and transverse photon flux is given by: $\varepsilon=\frac{1-y-\gamma^2y^2/4}{1-y+y^2/2+\gamma^2y^2/4}$.
The structure function $F^{\sin \ph}_{LU}$  receives contributions 
from the convolution of twist-2 and twist-3 distribution and fragmentation functions,
such as the twist-2 Boer-Mulders DF $h_1^\perp$ (\cite{Boer:1997nt,Pasquini:2010af}), the Collins FF $H_1^{\perp}$,
and the twist-3 DFs $e$ and $g^\perp$.
The Boer-Mulders DF $h_1^\perp$ describes the correlation between the transverse motion
of a quark and its own transverse spin, while $g^{\perp}$ can be interpreted as a higher twist analog of the Sivers function. Both functions represent spin-orbit correlations. 
The functions $\frac{\tilde{G^\perp }}{z}=\frac{G^\perp }{z}-\frac{m_q}{M_h}H_1^{\perp}$ and 
$\frac{\tilde{E}}{z}=\frac{E}{z}-\frac{m_q}{M_h}D_1$ are
interaction-dependent parts of the higher-twist FFs $G^\perp$ and $E$, respectively, in which $m_q$ is the quark mass. 
The quantities $f_1$ and $D_1$ are the usual unpolarized twist-2 DF and FF, respectively. 

The structure function $F^{\sin \ph}_{LU}$ in Eq.~\ref{FLUf} is higher-twist by nature. 
Thus, related observables such as beam-spin asymmetries in single-pion production off an 
unpolarized target can only be accessed at moderate values of $Q^2$.
Such higher-twist observables are a key for understanding
long-range quark-gluon dynamics.
They have also been interpreted in terms of 
average transverse forces acting on a quark at the instant after
absorbing the virtual photon \cite{Burkardt:2008vd}.

Different contributions to the structure function in Eq.~\ref{FLUf} have been 
calculated, related to both internal quark motion and the Collins mechanisms.
Sizable beam SSAs were predicted for pion production \cite{Yuan:2003gu} with spin-orbit correlations  
as the dynamical origin. Within this framework, the asymmetry generated at the
distribution level is given by either the convolution of the T-odd Boer-Mulders DF
$h_1^{\perp}$ with the twist-3 FF $E$ 
\cite{Jaffe:1991ra}, or the convolution of the  twist-3 T-odd DF
$g^\perp$ with the unpolarized FF $D_1$\cite{Metz:2004je}.

In contrast, calculations based on the Collins mechanism, $eH^{\perp}_{1}$,
predict vanishing beam SSAs for neutral pions
\cite{Schweitzer:2003uy,Efremov:2006qm,SchweitzerPrivate}.
The surprising characteristic that favored and unfavored Collins FFs are 
roughly equal in magnitude but opposite in sign, as
indicated by the latest measurements from HERMES \cite{Airapetian:2010ds},
COMPASS \cite{Alekseev:2010rw} and Belle \cite{Seidl:2008xc},
put the $\pi^0$ in a unique position in SSA studies
since the $\pi^0$ FF is the average of $\pi^+$ and $\pi^-$ FFs. 
Contributions to the beam SSA related to spin-orbit correlations could thus be studied without 
a significant background from the Collins mechanism.

Measurements of  beam-spin asymmetries in the electroproduction of
neutral pions in deep-inelastic scattering are presented from the E01-113 CLAS data set 
using a 5.776 GeV electron beam and the CEBAF Large Acceptance 
Spectro\-meter (CLAS) \cite{Mecking:2003zu} at Jefferson Laboratory.
Longitudinally polarized electrons were scattered off
an unpolarized liquid-hydrogen target. 
The beam polarization was frequently measured  with a 
M{\o}ller polarimeter 
and the beam helicity was flipped every 30 ms to minimize systematic instrumental effects.
Scattered electrons were detected in CLAS.
Electron candidates were selected by a hardware trigger using a 
coincidence of the gas Cherenkov counters and the lead-scintillator electromagnetic calorimeters (EC). 

\begin{figure}[h]
%\begin{minipage}{16pc}
\begin{center}
\includegraphics[height=.25\textheight,width=0.5\textwidth]{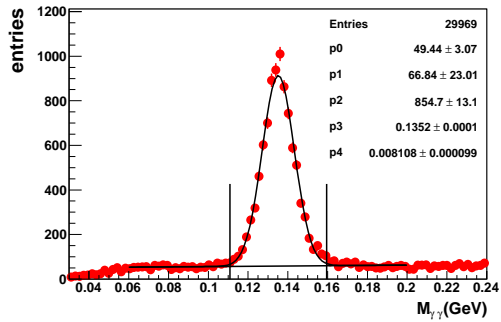}
\end{center}
\caption{\label{pi0bgkfit}(Color online) Invariant mass spectrum of the two photon ($\gamma \gamma$) system $M_{\gamma \gamma}$ in an arbitrarily chosen $\xbj$, $P_T$, $z$ and $\phi _{h}$-bin, fitted by a Gaussian plus a linear polynomial. Vertical black lines indicate $\pm 3\sigma$ from the mean.}
%\end{minipage}\hspace{2pc}%
%\begin{minipage}{16pc}
\end{figure}

Neutral pions were identified by calculating the invariant mass of two photons
detected with the CLAS EC and the Inner Calorimeter (IC) \cite{girod:2007jq}.
For events with more than two photons, the pair-wise combination of all photons was used.
In each kinematic bin, $\pi^0$ events were selected by a Gaussian plus linear polynomial fit to the two-photon invariant mass distribution (see Fig.\ref{pi0bgkfit}). 
In each $\phi_h$ bin and for each beam helicity, the combinatorial background was subtracted 
using the linear component of the fit, and $\pi^0$ events were selected within 
the invariant mass region defined by the mean of the Gaussian  $\pm 3 \sigma$, 
as indicated by the vertical lines in Fig.\ref{pi0bgkfit}.

\begin{figure}[h]
\begin{center}
\includegraphics[height=.25\textheight,width=0.5\textwidth]{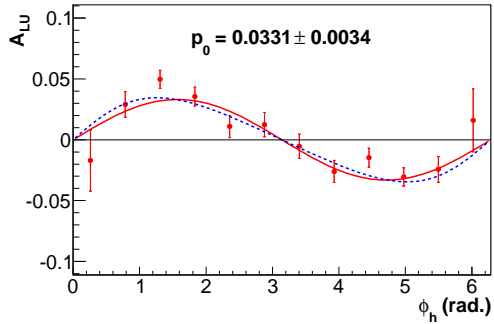}
\end{center}
\caption{\label{exFITsin12bin}(Color online) Examples of fits to the $A_{LU}$ asymmetry for $0.4<z<0.7$, $0.1<\xbj<0.2$ and $0.2$ GeV$ <P_T<0.4$ GeV using $p_0  \sin \ph $ (solid line) and  $p_0  \sin \ph /(1+p_1 \cos \ph)$ (dashed line). Both fits yield consistent amplitudes and $\chi^2$ per degree of freedom ($p_0=0.0331\pm0.0034$, $\chi^2$/ndf$ = 1.387$ and $p_0=0.0329\pm0.0034$, $\chi^2/$ndf$ = 1.31$, respectively). Only statistical error bars are shown.}
%\end{minipage} 
\end{figure}

\begin{figure*}[ht!]
\begin{center}
\includegraphics[width=0.7\textwidth]{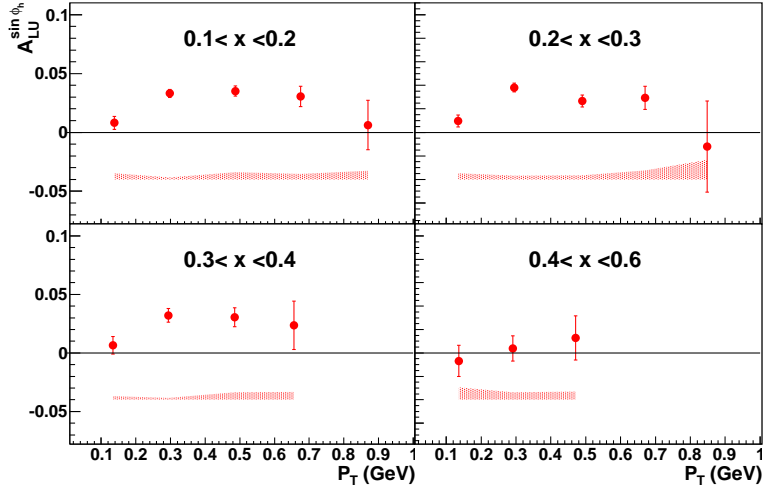}
\end{center}
\caption{(Color online) Asymmetry moment $A^{\sin \phi_h}_{LU}$ versus $P_T$ for different $\xbj$ ranges
and $0.4<z<0.7$. The error bars correspond to statistical and the bands to systematic uncertainties.
An additional 3\% scaling uncertainty arises from the beam polarization measurement and another 3\% relative uncertainty 
from radiative effects which are not included in the band.} 
 \label{ALU_vs_PTXB_sysH}
\end{figure*}
\begin{figure*}[ht!]
\begin{center}
\includegraphics[width=0.7\textwidth]{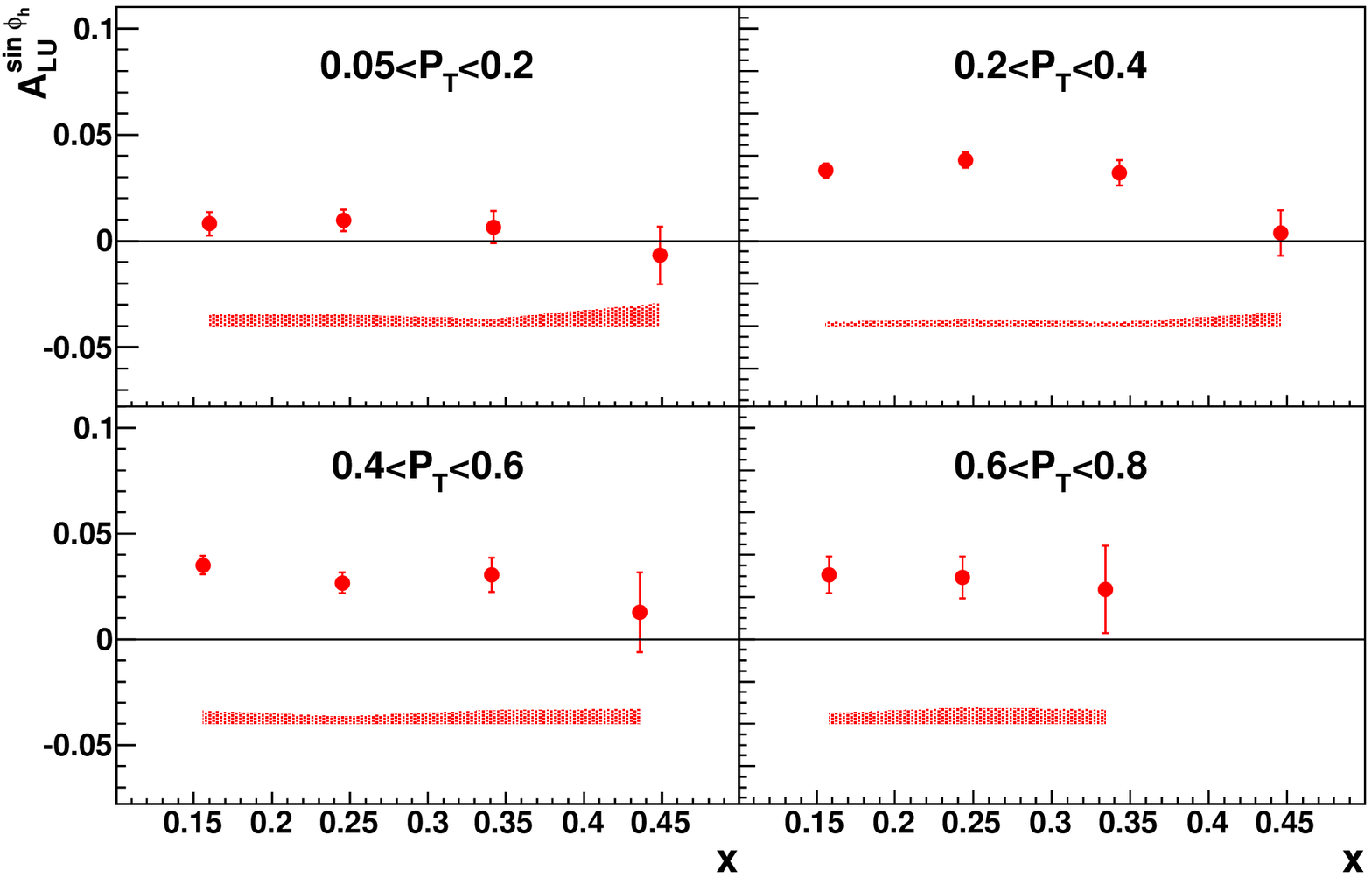}
\end{center}
\caption{(Color online) Asymmetry moment $A^{\sin \phi_h}_{LU}$ versus $\xbj$ for different $P_T$ ranges
and $0.4<z<0.7$. The error bars correspond to statistical and the bands to systematic uncertainties.
Complementary plot of Fig.~\ref{ALU_vs_PTXB_sysH}.} 
 \label{ALU_vs_XBfixPT_hod}
\end{figure*}

Deep-inelastic scattering events were selected by requiring $Q^2>1$ GeV$^2$ and  $W^2>4$ GeV$^2$,
where $W$ is the invariant mass of the hadronic final state. Events with missing-mass values for the $e\pi^0$ system that are smaller than 1.5 GeV ($M_{x}(e\pi^0)<1.5$ GeV) were discarded to exclude contributions from exclusive processes. 
A minimum value for the $\pi^0$ transverse momentum, $P_T>0.05$ GeV, ensures that the azimuthal
angle $\ph$ is well-defined. The total number of selected $e\pi^0$ coincidences  
was $\approx 3.0\times 10^6$ for the presented $z$ range, $0.4<z<0.7$, which selects the semi-inclusive region \cite{Avakian:2010ae}.

The beam-spin asymmetry $A_{LU} (\phi_{h})$ has been calculated for each kinematic bin as:
\begin{eqnarray}
A_{LU} (\phi_{h})=\frac{1}{P} \frac{N^{+}_{\pi^0}(\phi_{h}) - N^{-}_{\pi^0}(\phi_{h})}{N^{+}_{\pi^0}(\phi_{h}) + N^{-}_{\pi^0}(\phi_{h}) },
\label{alueq0}
\end{eqnarray}
where $P=0.794 \pm 0.024$ is the absolute beam polarization for this data set and $N^{+}_{\pi^0}$ and $ N^{-}_{\pi^0}$ are the number of $\pi^0$'s for positive and negative beam helicity, normalized to the respective integrated charges. The number of $\pi^0$'s is estimated by the integral of the histogram in the $\pm 3 \sigma$ range, minus the integral of the linear component of the fit.
Asymmetry moments were extracted by
fitting the $\ph$-distribution of $A_{LU}$ in each $\xbj$ and $P_T$ bin with the theoretically motivated 
function $p_0 \sin \ph $.
An example of this fit is shown in Fig.~\ref{exFITsin12bin} for a representative kinematic bin. 

In Fig.~\ref{ALU_vs_PTXB_sysH}, the
extracted $A_{LU}^{sin\phi}$ moment  is presented as a function of $P_T$ for different $\xbj$ ranges. The results are summarized in Table \ref{aluxbtb}.
Systematic uncertainties, represented by the bands at the bottom of
each panel, include the uncertainties due to the background subtraction,
the event selection and possible contributions of higher
harmonics. The first two contributions were estimated as the difference between 
the asymmetry moment extracted from data sets obtained with 
or without background subtraction, and by selecting the $\pi^0$ from the 
combination of all photons in an event or from events with exactly two 
photons. The contribution of higher harmonics
was estimated by employing 
the fit functions $p_0 \sin \ph $ or $p_0 
\sin \phi_h / (1+p_1  \cos \phi_h) $. The contributions from other 
harmonics such as $\sin 2\phi_h $ or $\cos 2\phi_h $ were also tested and 
found to be negligible. All the above contributions were added in quadrature.

An additional 3\% scaling uncertainty due to the beam polarization
measurements should be added to the above-mentioned systematic  
uncertainties. Radiative corrections have not been applied. However they have been
estimated to be negligible for the $\sin \ph$ modulation \cite{Avakian:2010ae,Afanasprivate} with an overall relative accuracy of 3\%.

The $A^{\sin \phi_h}_{LU}$ moment increases with increasing $P_T$ and reaches a maximum at $P_T \approx 0.4$ GeV.
There is an indication, within the available uncertainties, that the expected decrease of $A^{\sin \phi_h}_{LU}$ at larger $P_T$
could start already at $P_T \approx 0.7$ GeV. 
As a function of $\xbj$, $A^{\sin \phi_h}_{LU}$ appears to be flat in all $P_T$ ranges shown in Fig.~\ref{ALU_vs_XBfixPT_hod}. Note, however, that $Q^2$ varies with $\xbj$ (see Table~\ref{aluxbtb}).

\begin{figure}[h]
\begin{center}
\includegraphics[height=.25\textheight,width=0.5\textwidth]{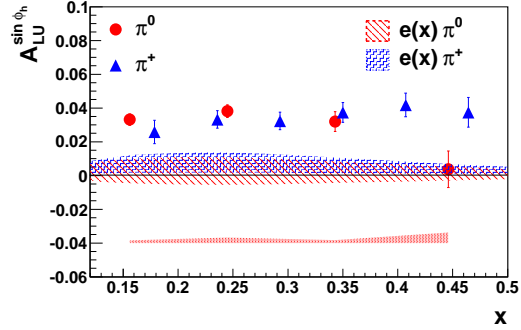}
\end{center}
\caption{(Color online) The $\pi^0$ beam-spin asymmetry moment $A^{\sin \phi_h}_{LU}$ 
vs. $\xbj$ compared to that of $\pi^+$ from an earlier CLAS measurement~\cite{HAdubna}.
Uncertainties are displayed as in Fig.~\ref{ALU_vs_PTXB_sysH}.
For both data sets $<P_T>\approx 0.38$ GeV and $0.4<z<0.7$. The right-hatched and left-hatched bands are model calculations involving solely the contribution from the Collins-effect \cite{SchweitzerPrivate}.}
\label{CLASaul}
\end{figure}

The measured beam-spin asymmetry moment for $\pi^0$ appears to be comparable  with the $\pi ^+$ asymmetry 
from a former CLAS data set \cite{HAdubna} both in magnitude and sign, as shown in Fig.~\ref{CLASaul}. 
For both data sets the average $P_T$ is about 0.38 GeV. 
Also shown are model calculations of $A^{\sin \phi_h}_{LU}$, as indicated in the figure (right-hatched and left-hatched bands), which take only the contribution from Collins-effect $eH^{\perp}_{1}$ into account \cite{Schweitzer:2003uy,Efremov:2006qm,SchweitzerPrivate,Belitsky:1997zw}, suggesting that contributions from the Collins mechanism cannot be the dominant ones.
In contrast, preliminary calculations of $A^{\sin \phi_h}_{LU}$ for pions \cite{GambergPrivate}, based on the models from Refs.~\cite{Bacchetta:2007wc,Gamberg:2007wm}, demonstrate a 
non-zero contribution from $g^\perp$.
 Because this DF 
can be interpreted as the higher-twist analog of the Sivers function, 
it underscores the potential of beam SSAs for studying 
spin-orbit correlations. 

\begin{figure}[h!]
\begin{center}
\includegraphics[height=.25\textheight,width=0.5\textwidth]{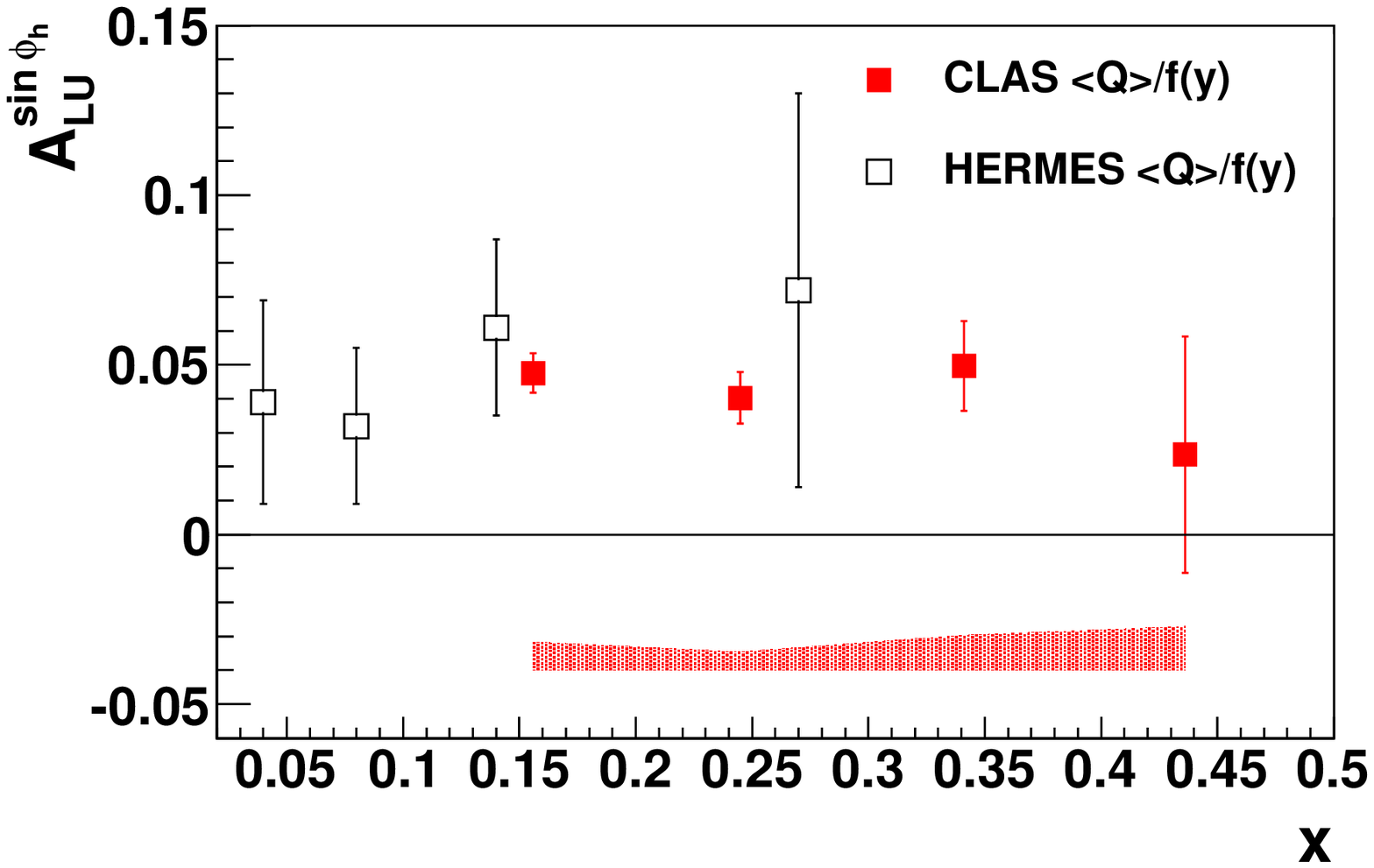}
\end{center}
\caption{\label{xbCH} (Color online) Asymmetry moment $A^{\sin \phi_h}_{LU}$ for $\pi^0$ multiplied  by the kinematic factor $<Q>/f(y)$ versus $\xbj$ from CLAS and HERMES \cite{Airapetian:2006rx}. The $0.4 <P_T<0.6$ GeV range of the CLAS data is used to compare with HERMES, because this yields average kinematics closest to HERMES.}
\end{figure}

\begin{figure}[h!]
\begin{center}
\includegraphics[height=.25\textheight,width=0.5\textwidth]{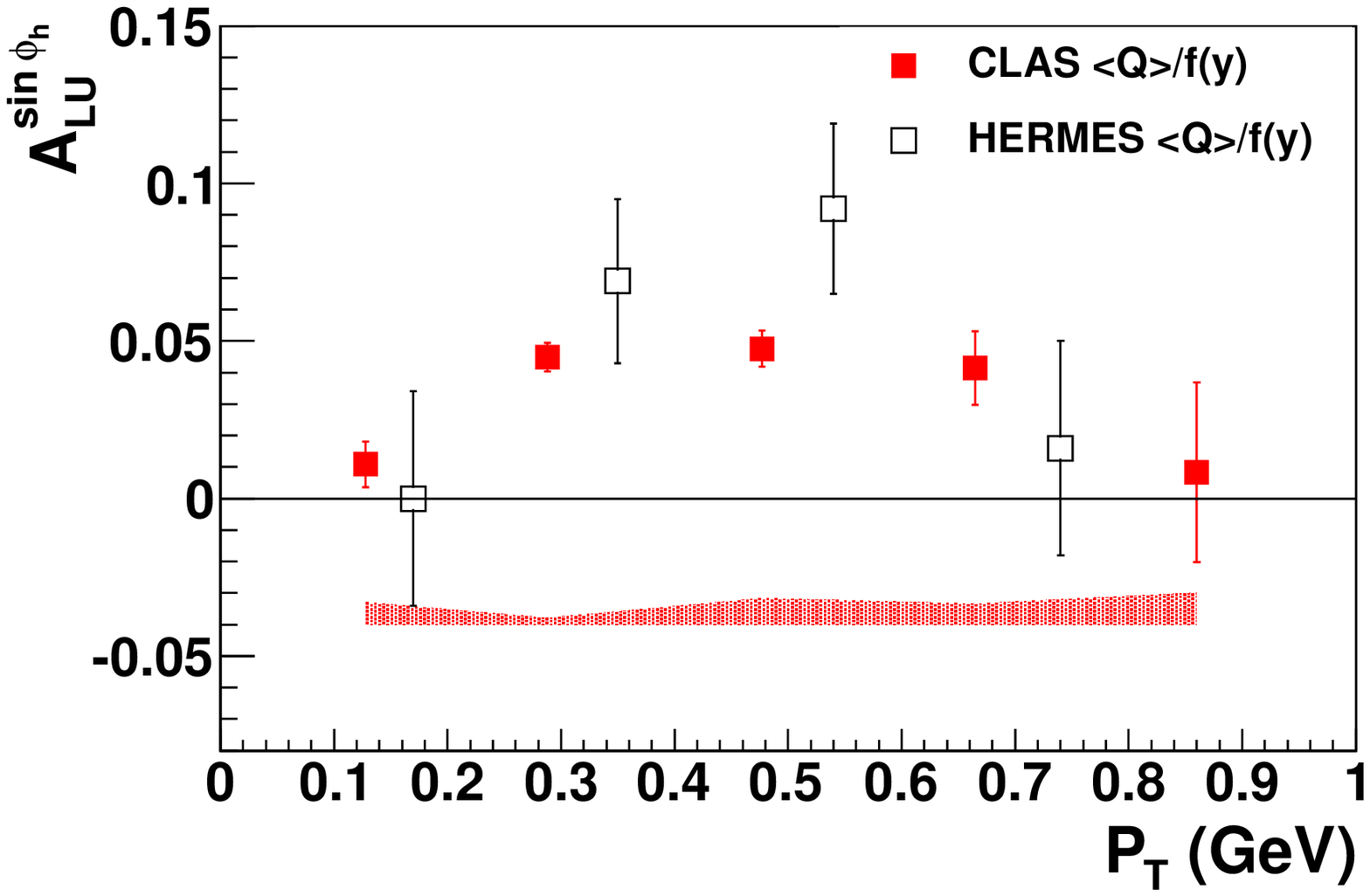}
\end{center}
\caption{\label{ptCH} (Color online) Asymmetry moment $A^{\sin \phi_h}_{LU}$ for $\pi^0$ multiplied by the kinematic factor $<Q>/f(y)$ versus $P_T$  from CLAS and HERMES \cite{Airapetian:2006rx} (the same as in Fig.~\ref{xbCH}). The $0.1<\xbj<0.2$ range of the CLAS data is used to compare with HERMES, as this yields average kinematics closest to HERMES.}
%\end{minipage} 
\end{figure}

Beam SSAs for charged and neutral pions were also measured
by the HERMES collaboration at a higher beam 
energy of 27.6 GeV~\cite{Airapetian:2006rx}. 
After taking into account the kinematic factors in the expression
of the beam-helicity-dependent and independent terms (\cite{Bacchetta:2006tn})
\begin{eqnarray}
f(y)=\frac{y \sqrt{1-y}}{1-y+y^{2}/2} ,
\label{fy}
\end{eqnarray}
 CLAS and HERMES measurements are found to 
be consistent with each other as shown in Figs.~\ref{xbCH} and~\ref{ptCH}, indicating that at energies as low as 4-6 GeV, the
behavior of beam spin asymmetries is similar to higher energy measurements. 
For comparison, CLAS data in the range $0.4$ GeV$ <P_T<0.6$ GeV are used in Fig.~\ref{xbCH} and in the range $0.1<\xbj<0.2$ in  Fig.~\ref{ptCH}, because these ranges yield average kinematic values similar to HERMES.

The CLAS data provide
significant improvements in the precision of beam SSA measurements for the kinematic region where the two data
sets overlap, and they extend the measurements to the large $\xbj$ region not accessible at HERMES.

%\section{Conclusions}

In summary, we have presented  measurements of the kinematic dependences
 of the beam-spin asymmetry in semi-inclusive $\pi ^0$ electroproduction from the E01-113 CLAS data set. 
The $\sin \phi_h$ amplitude was extracted as a function of $\xbj$ and transverse pion 
momentum $P_T$, for  $0.4<z<0.7$.
The asymmetry moment shows no significant $\xbj$ dependence for fixed $P_T$. Note, however, that $Q^2$ varies with $\xbj$ (see Table~\ref{aluxbtb}).
The observed asymmetry moment for $\pi ^0$ suggests that the major contribution to the pion beam SSAs 
originate from spin-orbit correlations. 

The results are compared with published HERMES data~\cite{Airapetian:2006rx}. They provide a significant improvement in precision and an important input for studies of higher-twist effects. 
Measured beam SSA's are in good agreement, both in magnitude and kinematic dependences, with measurements at significantly higher energies \cite{Airapetian:2006rx,Collaboration:2010fi}.

We thank A. Afanasev, A. Bacchetta, L. Gamberg, A. Kotzinian,  A. Prokudin, A. Metz and F.Yuan for useful and stimulating discussions.
We would like to acknowledge the outstanding efforts of the staff of the 
Accelerator and the Physics Divisions at JLab that made this experiment possible.
This work was supported in part by 
the National Science Foundation, 
the Italian Istituto Nazionale di Fisica Nucleare, 
the French Centre National de la Recherche Scientifique,
the French Commissariat \`{a} l'Energie Atomique, 
the National Reseach Foundation of Korea,
the UK Science and Technology Facilities Council (STFC),
the EU FP6 (HadronPhysics2, Grant Agreement number 227431),
the Physics Department at Moscow State University
and Chile grant FONDECYT N 1100872.
The Jefferson Science Associates (JSA) operates the Thomas Jefferson National Accelerator Facility for the United States Department of Energy under contract DE-AC05-06OR23177.

\begin{table*}[ht]
  \begin{center}	
    \begin{tabular}{|c|c|c|c|c||c|c|c|}
       \hline	
	  {$<P_T>$} & {$<z>$} & {$<\xbj >$} & {$<Q^2 >$} & {$<y>$} & {$A^{\sin \phi_h}_{LU}$} & {$\pm stat.$}  & {$\pm syst.$}\\
	  \hline 
	    \hline 
{0.138 } &  {0.507} & {0.160} & {1.36} & {0.786} & {0.0081} &  {0.0054} & {0.0053}\\
\hline
{0.298 } &  {0.517} & {0.156} & {1.35} & {0.797} & {0.0331} &  {0.0034} & {0.0016}\\
\hline
{0.487 } &  {0.528} & {0.156} & {1.34} & {0.798} & {0.0351} &  {0.0043} & {0.0061}\\
\hline
{0.675 } &  {0.553} & {0.158} & {1.36} & {0.795} & {0.0306} &  {0.0087} & {0.0048}\\
\hline
{0.870 } &  {0.513} & {0.154} & {1.34} & {0.800} & {0.0062} &  {0.0210} & {0.0074}\\
\hline
\hline
{0.134 } &  {0.515} & {0.246} & {1.97} & {0.739} & {0.0097} &  {0.0051} & {0.0054}\\
\hline
{0.295 } &  {0.521} & {0.245} & {1.98} & {0.747} & {0.0381} &  {0.0037} & {0.0033}\\
\hline
{0.490 } &  {0.516} & {0.245} & {1.97} & {0.745} & {0.0267} &  {0.0050} & {0.0036}\\
\hline
{0.670 } &  {0.517} & {0.243} & {1.97} & {0.752} & {0.0293} &  {0.0098} & {0.0076}\\
\hline
{0.848 } &  {0.484} & {0.233} & {1.99} & {0.788} & {-0.0121} &  {0.0386} & {0.0165}\\
\hline
\hline
{0.134 } &  {0.514} & {0.342} & {2.59} & {0.697} & {0.0066} &  {0.0075} & {0.0032}\\
\hline
{0.294 } &  {0.509} & {0.343} & {2.55} & {0.685} & {0.0320} &  {0.0059} & {0.0017}\\
\hline
{0.485 } &  {0.488} & {0.341} & {2.54} & {0.689} & {0.0305} &  {0.0081} & {0.0063}\\
\hline
{0.656 } &  {0.477} & {0.334} & {2.66} & {0.734} & {0.0236} &  {0.0208} & {0.0068}\\
\hline
\hline
{0.136 } &  {0.491} & {0.449} & {3.29} & {0.676} & {-0.0068} &  {0.0134} & {0.0106}\\
\hline
{0.291 } &  {0.478} & {0.446} & {3.21} & {0.661} & {0.0038} &  {0.0108} & {0.0063}\\
\hline
{0.471 } &  {0.457} & {0.436} & {3.26} & {0.690} & {0.0128} &  {0.0189} & {0.0069}\\
	  \hline
    \end{tabular}	
    \caption{The asymmetry moments $A^{\sin \phi_h}_{LU}$ and their statistical and systematic uncertainties at average values of $P_T$, $z$, $\xbj$, $Q^2$ and $y$. An additional 3\% scaling uncertainty from the beam polarization measurement and another 3\% relative uncertainty from radiative effects should be added to the total uncertainty.} 
    \label{aluxbtb}
  \end{center}	
\end{table*}
\bibliographystyle{model1a-num-names}
\bibliography{alu}
%\bibliography{alu}

\end{document}